\documentclass[a4paper,12pt]{article}
\usepackage{amsmath,amssymb,amsthm}
\usepackage[width=17cm,height=23cm]{geometry}
\usepackage{graphicx}

\newcommand{\tmop}[1]{\ensuremath{\operatorname{#1}}}

\theoremstyle{plain}
\newtheorem{definition}{Definition}
\newtheorem{lemma}[definition]{Lemma}
\newtheorem{theorem}[definition]{Theorem}

\newtheorem{proposition}[definition]{Proposition}

\theoremstyle{remark}

\newtheorem{remark}[definition]{Remark}

\begin{document}

\title{Spectral statistics along sequences of irreducible representations }
\author{Ingolf Sch\"afer \\
\textit{Fakult\"at f\"ur Mathematik, Ruhr-Universit\"at Bochum, D-44780 Bochum,
Germany} \\
\\
Marek Ku\'s \\
\textit{Centrum Fizyki Teoretycznej PAN, Al. Lotnik\'ow 32/42, 02-668 Warszawa,
Poland}}
\date{\today}
\maketitle

\section{Introduction}
\label{sec:introduction}

Quantum systems in finite dimensional Hilbert spaces proved to be a good
laboratory for investigations of connections between quantum and classical
characteristics of chaotic and integrable dynamics. In particular statistical
properties of quantum spectra are well established signatures of integrability
on the classical level \cite{haake00}.

Recently we investigated a general procedure of attaining the classical limit
for quantum systems with Hamiltonians defined as polynomials in generators of
finite-dimensional compact Lie algebras \cite{schaeferkus}. We generalized to
an arbitrary algebra $\mathfrak{g}$ the procedure proposed in
\cite{gnutzmannkus} and \cite{haakekus}, where a particular example of the
$\mathfrak{su}_3$ algebra was used to illustrate connection between spectral
statistics and properties of the limiting classical Hamiltonian system. In the
above cited papers we discuss at length the physical relevance of model
Hamiltonians defined in terms of generators of an arbitrary $\mathfrak{g}$, as
well as the meaning of (various) classical limits of them, so here we only
mention that they describe e.g., collections of many (say $N$) identical
multilevel (or spin-like) systems. The classical limit is attained when $N$
goes to infinity and is performed by increasing to infinity the dimension of
the irreducible representation in the space of which the Hamiltonian of the
system acts. In cases when the rank of the Lie algebra is greater than one, no
unique way of attaining the classical limit via a sequence of irreducible
representation with increasing dimensionality can be, \textit{a priori},
singled out. This fact has some interesting consequences discussed thoroughly
in the above cited papers (like e.g., different limiting statistics of the
eigenvalues of the same system, but for different paths to the classical
limit).

In the present paper we want to study more closely spectral statistics of
Hamiltonians defined in terms of elements of $\mathfrak{su}_3$ along sequences
of irreducible representations mimicking the transition to the classical limit.
Formally, the classical limit, is often treated as `going with the value of the
Planck constant to zero'. Such a statement taken literally is meaningless - we
should specify with respect to value of which physical quantity the Planck
constant becomes negligible, so that the classical description of the system is
justified. In the above mentioned situation of $N$ $M$-level systems and a
`nonlinear' Hamiltonian in the form of a polynomial in the generators of
$\mathfrak{su}_n$ the classical limit was obtained for $N\to\infty$, and the
`effective Planck constant' (the quantity which approached zero in the
classical limit) was scaled appropriately with the dimensions of
representations. Such a scaling was motivated by purely physical
considerations. In the present paper we show that for an arbitrary
$\mathfrak{g}$ there is a unique proper scaling relevant for the classical
limit. This is one of the main result of the paper justifying our approach to
study the outlined class of models in order to understand the transition to
classical limit in well controllable models.

In the following we will concentrate on Hamiltonians taking the form of second
order polynomial in elements of $\mathfrak{su}_3$, inspired by the so called
Lipkin Hamiltonian of a three-shell nuclear model, often invoked in various
investigation of quantum-chaotic phenomena \cite{leboeuf90a,leboeuf90b,wang98}.
We will be able to thoroughly investigate limiting spectral properties of the
linear part of the Hamiltonian in all possible cases of attaining the classical
limit and give some remarks on the properties of the full nonlinear
Hamiltonian, which will be an object of our further studies in a forthcoming
paper.

Let us briefly sketch the contents of the remaining sections. In the following
one we give the definition of the nearest neighbour distribution of eigenvalues
of a hermitian matrix - the main object characterizing a quantum system
investigated in the paper. Section \ref{sec:setting} is concerned with rays of
irreducible representations and an abstract notion of hermitian operator. In
Section \ref{sec:simple-operators} a limit theorem for simple operators, i.e.,
Lie algebra elements is given. Under some assumptions on the starting
representation these operators will have Dirac statistics as limiting nearest
neighbour distribution. A proof of this is given by estimating the number of
different eigenvalues of the represented operator along the ray through the
starting representation. As a corollary we obtain Dirac statistics for certain
monomials in $\mathcal{U}(\mathfrak{g})$.

If we consider operators which are not given as monomials there is need for a
suitable choice of rescaling.  The already announced main result of Section
\ref{sec:rescaling} is that under all possible ways of rescaling only the one
used in \cite{haakekus} and \cite{schaeferkus} gives interesting and physically
meaningful results. For this we give physical arguments as well as mathematical
facts.

In the last section we deal with an explicit 'nonlinear' example. We consider
the full Lipkin-Hamiltonian. By giving explicit formulas for the action of
operators in arbitrary representation we are able to complete our statements
from Section \ref{sec:simple-operators} concerning the nearest neighbour
statistics for the linear part for arbitrary choice of the sequence of
representations and make some remarks about the structure of the matrix of the
full Lipkin Hamiltonian in arbitrary representation.

\section{Nearest neighbour distributions}

The statistics we deal with are the nearest neighbour distributions which are
defined as follows

\begin{definition}
  Let $A$ be an $N\times N$ hermitian matrix with eigenvalues, counted
  with multiplicity, $x_1\leq\dots\leq x_N$.

  The \textbf{nearest neighbour distribution} of $A$ is the Borel
  measure on $\mathbb{R}$ given by
  \begin{equation}
    \mu_A = \frac{1}{N}\sum_{i=1}^{N-1}
    \delta_{\frac{N}{x_N-x_1}\cdot(x_{j+1}-x_j)}
  \end{equation}
  if $x_1\neq x_N$, and
  \begin{equation}
    \mu_A = \frac{N-1}{N}\delta_0
  \end{equation}
  if $x_1=\ldots=x_N$.
\end{definition}

In order to have a notion of limit of these distribution we make use of the
following distance between measures which is common in statistics
\begin{definition}
  Let $\mu,\nu$ be Borel measures on $\mathbb{R}_{\geq 0}$ of finite
  mass.  The \textbf{Kolmogorov-Smirnov} distance of $\mu$ and $\nu$
  is given by
  \begin{equation}
    d_{KS}(\mu,\nu):=\sup_{x\in\mathbb{R}_{\geq 0}}\left|\int_0^x
      d\mu-\int_0^xd\nu \right|.
  \end{equation}
\end{definition}

We will use the notion of convergence induced by this distance exclusively in
text. Since weak convergence of measure is implied by the pointwise convergence
of distribution functions the convergence in the sense of Kolmogorov-Smirnov
also implies weak convergence.

\section{Setting}
\label{sec:setting}

We fix a semi-simple, compact Lie group $K$ and its complefixication $G$ whose
Lie algebras are denoted by $\mathfrak{k}$ and $\mathfrak{g}$. We denote the
universal enveloping algebra of $\mathfrak{g}$ by $\mathcal{U}(\mathfrak{g})$.

By convention each considered representation of $K$ will be irreducible and
unitary and therefore finite-dimensional. To have a simple notation we use the
same letter for the representations of $K$ and $G$ and the induced Lie algebra
representation. Moreover, this letter will be used for the induced
representation on $\mathcal{U}(\mathfrak{g})$. If $\rho: K \to \tmop{U}(V)$ is
such an irreducible, unitary representation, $\rho:
\mathcal{U}(\mathfrak{g})\to \tmop{End}(V)$ is thus the induced representation
on the universal enveloping algebra from the original $\rho$.

There are numerous possibilities of sequences of irreducible representations of
$K$, but using the results from \cite{schaeferkus} a physical meaningful limit
is provided by \emph{rays} of irreducible representations.

\begin{definition}
  Let $\rho:K \to \tmop{U}(V)$ be
  an irreducible representation of highest weight $\lambda$.

  A \textbf{ray} through $\rho$ is a sequence of irreducible
  representations $(\rho_k:\mathcal{U}(\mathfrak{g}) \to
  End(V_k))_{k\in\mathbb{N}}$ , such that $\rho_k$ has the highest
  weight $k\cdot\lambda$.
\end{definition}

We would like to emphasize that the irreducible representations of highest
weight $k\cdot\lambda$ are exactly the leading irreducible representations in
the $k$-fold tensor product of the representation $\rho$.

Since we would like to treat the operators to be discussed as Hamiltonians of
some physical systems, we need a representation independent notion of hermitian
operators.

\begin{lemma}
  There exists a unique $\mathbb{R}$-linear map
  $\dagger:\mathcal{U}(\mathfrak{g})\to \mathcal{U}(\mathfrak{g})$
  satisfying
  \begin{itemize}
  \item $\xi=-\xi^{\dagger}$ for $\xi\in \mathfrak{k}$
  \item $(c\xi_1\dots\xi_p)^{\dagger}=\bar{c}\xi^{\dagger}_p\dots
    \xi_1^{\dagger}$ for $\xi_j\in\mathcal{U}(\mathfrak{g}),c\in\mathbb{C}$
  \end{itemize}
  and for all $\xi\in \mathcal{U}(\mathfrak{g})$ with
  \begin{equation}
    \xi^{\dagger}=\xi
  \end{equation}
  $\rho(\xi)$ is hermitian, for each irreducible, unitary
  representation $\rho:K \to \tmop{End}(V)$.
\end{lemma}

\begin{proof}
  First, one defines $^\dagger$ on the full tensor algebra
  $\mathcal{T}(\mathfrak{g})$ by the above requirements and checks by a direct
  calculation that is passes to the quotient.

  The statement about the image $\rho(\xi)$ follows by another calculation from
  the defining properties and the fact that $\rho:\mathcal{U} (\mathfrak{g})
  \to \tmop{End}(V)$ is an algebra homomorphism.
\end{proof}

Thus, we may propose the following definition.
\begin{definition}
  An operator $\xi \in \mathcal{U}(\mathfrak{g})$ is called abstract
  hermitian operator, if $\xi=\xi^\dagger$. The real subspace of
  abstract hermitian operators in $\mathcal{U}(\mathfrak{g})$ is
  denoted by $\mathcal{H}$.
\end{definition}

\begin{remark}
  Note, that $\xi$ may not be an abstract hermitian operator although
  $\rho(\xi)$ is hermitian, e.g., every $\xi$ in the kernel of $\rho$
  is represented as hermitian matrix.
\end{remark}

\section{Simple Operators}
\label{sec:simple-operators}
In this section the spectral statistics of simple operators, i.e.,
hermitian generators of a semi-simple compact Lie group, are
considered along sequences of irreducible representations.
The main result of this section is the following
\begin{theorem}
  \label{thm:simple-operators}
  If $G$ is simple, $\mathrm{rank}\ G\geq 2$ and $\rho$ is an irreducible
  representation of highest weight $\lambda$, such that $\lambda$ is
  in the interior of the Weyl chamber, then for all $\xi\in
  i\mathfrak{k}$
  \begin{equation}
      \lim_{k\to \infty} \mu_{\rho_k(\xi)} = \delta_0
  \end{equation}
\end{theorem}

The strategy for the proof is as follows. First, $\xi$ is contained in the
complexified Lie algebra $\mathfrak{t}^{\mathbb{C}}$ of some fixed maximal
torus $T$. We compare the dimension of the representation space $V_k$ with the
number of different weights of the representation, since the weights evaluated
at $\xi$ are the eigenvalues of $\rho_k(\xi)$. The quotient of these numbers is
shown to converge to zero, hence in the limit we will recover the Dirac measure
for the nearest neighbor distribution of $\rho_k(\xi)$.

We start by giving an estimation the dimension of an irreducible representation
in terms of the highest weight.
\begin{lemma}
\label{lem:estimations} Let $\rho$ be an irreducible representation with the
highest weight $\lambda$ and $\lambda=\sum_j \lambda_jf_j$ the decomposition of
$\lambda$ into the fundamental weights $f_j$.

Then the dimension of $\rho_\lambda$ is bounded from below by
\begin{equation}
  \label{eq:dim-estimate} \tmop{dim} \rho_\lambda \geq
  \prod_{\alpha\in\Pi^+,\langle\lambda,\alpha\rangle>0}
   \frac{\langle \lambda,
    \alpha \rangle}{\langle\delta,\alpha\rangle},
\end{equation}
where $\Pi^+$ denotes the set of positive roots and
$\delta=\frac{1}{2}\sum_{\alpha\in\Pi^+}\alpha$.

Moreover, the number $n_\lambda$ of possible weights of $\rho_\lambda$ is
bounded from above as follows
\begin{equation}
  \label{eq:weight-estimate}
  n_\lambda\ \leq\ \tmop{ord}(W)\cdot\prod_{j}(\lambda_j+1).
\end{equation}
\end{lemma}

\begin{proof}
  The Weyl's dimension formula \cite{wallachgoodman} reads
  \begin{equation}
  \tmop{dim} \rho_\lambda = \prod_{\alpha\in\Pi^+}\frac{\langle\delta+\lambda,
      \alpha \rangle}{\langle\delta,\alpha\rangle} = \prod_{\alpha\in\Pi^+}
      \left( 1 +\frac{\langle \lambda,
        \alpha \rangle}{\langle\delta,\alpha\rangle} \right).
  \end{equation}
  Now, $\langle \lambda,\alpha \rangle \geq 0$ and $\langle \delta, \alpha
  \rangle>0$ for all positive roots $\alpha$.  Thus, the first inequality is
  clear.

  Starting from $\lambda$ we get all other weights by subtracting multiples of
  the roots. The lattice of roots is a sublattice of the lattice of weights, so
  we can reach every weight by subtracting multiples of the fundamental weights
  $f_j$.

  There are at most $\prod_j (\lambda_j+1)$ of such possible substractions
  which give positive weights and every weight is in the $W$-orbit of a
  positive weight, which has at most $|W|$ elements. This proves the second
  inequality.
\end{proof}

Now, we give a proof of Theorem \ref{thm:simple-operators}.
\begin{proof}[Proof of Theorem \ref{thm:simple-operators}]
  Let $Q$ be the set
  \begin{equation}
    Q:=\{ \alpha\in\Pi^+ : \langle \lambda,\alpha\rangle > 0 \}
  \end{equation}
  and $q:=\tmop{card} Q$. We claim that $Q=\Pi_+$. Indeed, since $\lambda$ is
  in the interior of the Weyl chamber, we have
  \begin{equation}
    \langle \lambda,\alpha_j\rangle > 0
  \end{equation}
          for all simple roots $\alpha_j$. But the positive roots are just positive
  linear combinations of simple roots, thus $Q=\Pi_+$.

  There exist exactly $r:=\tmop{rank} G$ simple roots and $q>r$ because $G$ is
  assumed to be simple and $r\ge 2$. If $q=r$ were true, the root system could
  be decomposed into $1$-dimensional pieces, which would contradict the
  assumptions.

  Combining the above two estimates, (\ref{eq:dim-estimate}) and
  (\ref{eq:weight-estimate}), we obtain
  \begin{equation}
    \label{eq:mult-estimation}
    \frac{\#\text{differ. eigenvalues of
      }\rho_k(\xi)}{\tmop{dim}\rho_k} \leq
    \frac{\tmop{ord}(W)\prod_j (k\lambda_j+1)}{
      \prod_{\alpha\in Q}\frac{\langle k\lambda,\alpha\rangle}{
      \langle \delta,\alpha\rangle}} < \tmop{const}(\lambda) k^{r-q},
  \end{equation}
  since the number of different eigenvalues is less then the number of weights
  of the representation.

  The right-hand side of (\ref{eq:mult-estimation}) converges to zero if $k$
  goes to $\infty$ which finishes the proof.
\end{proof}

\begin{remark}
  Theorem \ref{thm:simple-operators} is still valid for $G$ semi-simple
  provided that $q>\tmop{rank} G$.
\end{remark}

\section{Rescaling}
\label{sec:rescaling}

If we consider operators $\xi\in\mathcal{H}$ which are not `homogeneous', e.g.,
$\xi=\eta_1+\eta_2\eta_3$, there is a question of the right scaling of the
operators when the dimensionality of irreducible representations increases. It
will in general happen that the distance between the maximal and minimal
eigenvalues of $\eta_2\eta_3$ along a ray of representations grow much faster
than the distance for $\eta_1$. Thus, as will be shown, the lower degree terms
vanish in the limit. It is therefore necessary to rescale the operators in
dependence of the parameter $k$.

\begin{definition}
  Fix a basis $\xi_1,\dots,\xi_n$ of $\mathfrak{g}$ and identify
  $\mathcal{U}(\mathfrak{g})$ with $\mathbb{C}[\xi_1,..,\xi_n]$. Let
  $\rho:K\to\tmop{U}(V)$ be an irreducible representation and
  $(\rho_k)_{k\in \mathbb{N}}$ be the ray through $\rho$.

  A rescaling map is a family of maps $(r_k:\mathcal{U}(\mathfrak{g})
  \to \mathcal{U}(\mathfrak{g}))_{k\in\mathbb{N}}$ given as
  \begin{equation}
    r_k(\xi_j) = \frac{1}{s_k}\xi_j
  \end{equation}
  on the generators and extended multiplicatively, where each
      $s_k\in\mathbb{N}^\ast$ and $\lim_{k\to\infty} s_k=\infty$.
\end{definition}

In other words we put the elements in $\mathcal{U}(\mathfrak{g})$ in
normal ordering with respect to the chosen basis and substitute
$\frac{1}{s_k}\xi_j$ for each $\xi_j$. Thus, rescaling is induced by
the linear automorphism of $\mathfrak{g}$ given by multiplication with
$s_k^{-1}$. Although this automorphism is independent of the chosen
basis, the resulting extension to a vector space automorphism of
$\mathcal{U}(\mathfrak{g})$ is not, since it depends on the
identification of $\mathcal{U}(\mathfrak{g})$ with
$\mathbb{C}[X_1,\dots,X_n]$. More precisely, it depends on the ordering
of basis vectors. Two different scaling procedures can be termed `natural'

%\noindent\textbf{Natural Examples.}
\begin{enumerate}
\item Rescaling by the dimension: $s_k:=\tmop{dim}\rho_k$
\item Rescaling by the parameter: $s_k:=k$ for $k>0$ and $s_0:=1$.
\end{enumerate}

Our main goal in this section is to show that only the latter will give
physically interesting results.

\subsection{A Physical Argument}
As indicated in the introduction and explained in details in
\cite{gnutzmannkus,haakekus,schaeferkus}, we want to treat the limiting
procedure described in the previous paragraphs as corresponding to taking the
classical limit of the physical system in question. Using the momentum map we
associate to an irreducible representation $\rho:K\to\tmop{U}(V)$ with highest
weight $\lambda$ the coadjoint orbit $K.\lambda$ in $\mathfrak{k}^\ast$. The
classical limit is the performed exactly by increasing $k$ to infinity along
the ray through $\rho$ the corresponding sequence of coadjoint orbits. In this
way $k\to\infty$ should correspond to $\hbar\to 0$.

We will show that $k$ can be identified with $\hbar^{-1}$ by invoking a
semiclassical argument to express the volume the phase space in terms of the
Planck constans. We will show that the rescaling by inverse parameter is the
only natural choice, since a first order operator should be measured in the
natural scale, i.e., in terms of $\hbar$, and their rescaling should be
\begin{equation}
  r_k(\xi_j)=\hbar \xi_j = \frac{1}{k}(\xi_j).
\end{equation}
Operators of higher degree should then be rescaled accordingly, which
implies rescaling by the parameter. To this end let us formulate
\begin{proposition}
  Let $\rho$ be an irreducible representation with highest weight
  $\lambda$ in the interior of the Weyl chamber. Then
  \begin{equation}
     \label{eq:hbar-scaling}
     k = \frac{1}{\hbar}.
  \end{equation}
\end{proposition}

\begin{proof}
For the proof of (\ref{eq:hbar-scaling}) recall that the volume of the
phase space should be proportional to
\begin{equation}
  \label{eq:proportionalityfactor}
  \frac{1}{\hbar^f},
\end{equation}
where $f$ denotes the number of degrees of freedom in the classical phase
space, i.e.\ the dimension of the coadjoint orbit through the highest weight.
By a standard fact of the theory of compact Lie groups \cite{} this number is equal to
the number of positive roots, i.e.,
\begin{equation}
  f = \ ^\#\Pi^+.
\end{equation}
But the volume of the phase space $K.\lambda$ can be expressed in
terms of the dimension of an irreducible representation. For this let
$\delta$ denote half of the sum of positive roots and set
$\lambda'=\lambda-\delta$. It is proved in \cite{kirillov68} that
\begin{equation}
  \tmop{vol}(K.\lambda)=\tmop{dim} \rho_{\lambda'}.
\end{equation}
Using Weyl's dimension formula it follows that
\begin{equation}
  \tmop{vol}(K.\lambda) = \prod_{\alpha\in\Pi^+}\frac{\langle
    \lambda'+\delta,\alpha\rangle}{\langle\delta,\alpha\rangle} =
  \prod_{\alpha\in\Pi^+}\frac{\langle \lambda,
   \alpha\rangle}{\langle\delta,\alpha\rangle}.
\end{equation}
Therefore the volume of $K.(k\lambda)$ is proportional to $k^{\
  ^\#\Pi^+}=k^f$. Comparing with the proportionality factor in
(\ref{eq:proportionalityfactor}) implies
\begin{equation}
  k = \frac{1}{\hbar}
\end{equation}
and completes the proof.
\end{proof}

\subsection{A Mathematical Argument}
\label{sec:math-argum}

Although the reader may be already convinced by the above reasoning, we would
like to give a more mathematical argument, which applies also to
representations on the border of the Weyl chamber.

We state the following lemma:
\begin{lemma}
  \label{lem:resc-spectr-stat}
  Let $\xi_H=\sum_{I} a_{I}\Xi^I \in \mathcal{U}(\mathfrak{g})$ be
  given with $\xi_H^\dagger=\xi_H$ and consider the ray
  $(\rho_k:K\to\tmop{U}(V_k))_{k\in\mathbb{N}^\ast}$ through an
  irreducible representation $\rho:K\to \tmop{U}(V)$ of highest weight
  $\lambda$.

  Then
  \begin{equation}
    \label{eq:resc-spectr-stat}
    \| \rho_k (\xi_H) \|_{\tmop{End}(V_k)} \leq
    \sum_{I} |a_{I}| c(\lambda)^{|I|} \cdot k^{|I|}
  \end{equation}
  where the $c_j(\lambda)$ is a constant depending only on $\lambda$ and
  $\|\cdot \|_{\tmop{End}(V_k)}$ denotes the operator norm on
  $\tmop{End}(V_k)$.
\end{lemma}

\begin{proof}
  We use the explicit construction of irreducible representations by
  Borel-Weil. For this let $r:=\tmop{rank}G$ and
  \begin{equation}
    \label{eq:tensor-generating-set}
       S_j = (s_1^{(j)}, \ldots, s_{d (j)}^{(j)}),\quad j = 1, \ldots, r
  \end{equation}
  denote a basis of the $j$-th fundamental representation. These are
  holomorphic sections in a holomorphic line bundle
  \begin{equation}
     L_j \rightarrow G / B
  \end{equation}
  where $B$ is a Borel subgroup of $G$ and $L = G \times_{\chi_{_j}}
  \mathbb{C}$, such that $\chi_j : B \rightarrow \mathbb{C}$ is the
  exponentiated character of the fundamental weight $\lambda_j$. The
  irreducible representation with highest weight $\lambda$ is then given by the
  action on sections of the line bundle
  \begin{equation}
     L = L_1^{\otimes \lambda_1} \otimes \ldots \otimes L_j^{\otimes \lambda_j}
     \rightarrow G / B .
  \end{equation}
  By the theorem of Borel-Weil the tensors of the form
  \begin{equation}
    \label{eq:tensors_in_representation}
    S_1^{I_1} \otimes \ldots \otimes S_r^{I_r},
  \end{equation}
  with $I_1, \ldots, I_r$ multiindices of degree $|I_j | = \lambda_j$
  constitute a generating system of the space of sections.

  Without loss of generality we may assume that $\xi_1$ is represented by a
  diagonal hermitian matrix in every fundamental representation, whose spectral
  norms a bounded by a constant $c_{\xi_1}$. Since the operator norm is equal
  to the spectral norm, we wish to give an estimate for the maximal absolute
  value of an eigenvalue of $\xi_1$ in $\rho_\lambda$.

  But on the generating system of vectors given by
  (\ref{eq:tensors_in_representation}) the action is on each
  factor separately, so we have
  \begin{equation}
      \| \rho(\xi_1)\| \leq c_{\xi_1}(\lambda_1 + \ldots + \lambda_r)
      =: c_{\xi_1}|\lambda|.
  \end{equation}
  For this recall, that $\xi_j$ acts as differential operator of degree 1,
  therefore we get the $\lambda_j$'s as scalars and not in the exponent.

  Clearly, the same argument can be carried out for $\xi_2, \ldots,
  \xi_n$. Thus, we have the following estimate
  \begin{equation}
    \| \rho_k (\xi_j)\| \leq ck (\lambda_1 + \ldots + \lambda_r) = ck|\lambda|
    \label{eigvestimate}
  \end{equation}
  for all $j = 1, \ldots, n$ with the constant $c:=\max_j c_{\xi_j}$.

  Now, consider $\gamma = \sum_I a_I X^I$. Then
  \begin{equation}
    \label{eq:operatornorm_estimation}
  \| \rho_k  (\gamma) \|_{\tmop{End} (V_k)}
     \leq \sum_I |a_I| \| \rho_k
     (\xi_1)\|_{\tmop{End} (V_k)}^{i_1} \cdot \ldots \cdot \| \rho_k
     (\xi_n)\|^{i_n}_{\tmop{End} (V_k)} .
  \end{equation}
  Using the estimates given by (\ref{eigvestimate}) and
  Lemma \ref{lem:estimations}, we see that
  \begin{equation}
    \label{eq:norm-estimate}
    \| \rho_k  (\gamma) \|_{\tmop{End} (V_k)}
    \leq \sum_I |a_I | k^{|I|} \cdot (c|\lambda|)^{|I|},
  \end{equation}
  which completes the proof.
\end{proof}

\begin{remark}
  If $\xi\in\mathfrak{g}$, $\xi^\dagger=\xi$ and $\rho_1(\xi)\neq
  0$, then the asymptotic growth of the norm is given by
  \begin{equation}
    \label{eq:asymptotic_norm}
    \|\rho_1(\xi)\| \sim c\cdot k,
  \end{equation}
  where $c$ is a constant, which depends on $\lambda$ and
  $\|\rho_1(\xi)\|$. This can be seen by a short calculation for the
  action of $\xi$ on the tensors in (\ref{eq:tensor-generating-set}).\ \\
\end{remark}
Recall that for a highest weight $\lambda$ the set $Q$ is
defined as $Q=\{\alpha\in\Pi_+\,:\,\langle \alpha,\lambda\rangle\}$
and $q=\,^\# Q$. We use the Lemma and the remark to prove the
following theorem
\begin{theorem}
  Consider the ray $(\rho_m:K\to\tmop{U}(V_m))_{m\in\mathbb{N}^\ast}$
  through an irreducible representation $\rho:K\to \tmop{U}(V)$ of
  highest weight $\lambda$ and assume $q>2$.

  Let $\xi_H=\xi+\sum_{|I|>1} a_{I}\Xi^I \in
  \mathcal{U}(\mathfrak{g})$ be given, such that $\xi_H^\dagger=\xi_H$
  and $\xi\in\mathfrak{g}$ with $\|\rho_1(\xi)\|\neq 0$. Then the
  nearest neighbor distributions of rescaled $\xi_H$ along the ray
  agree with the limit of the nearest neighbor distribution of $\xi$,
  if the rescaling factor $s_K$ grows at least like $k^{1+\epsilon}$.
\end{theorem}

\begin{proof}
  We claim that
  $\|\rho_k(r_k(\sum_{|I|>1})a_{I}\Xi^{I})\|_{\tmop{End}(V_k)}\to 0$, as
  $k\to\infty$. Making use of (\ref{eq:norm-estimate}) and the growth of $s_k$
  we obtain
  \begin{equation}
    \left\|\frac{1}{s_k^{|I|}}\rho(\Xi^I)\right\|_{\tmop{End}(V_k)} \leq
      \frac{k^{|I|}c|\lambda|^{|I|}}{s_k^{|I|}}\leq \tilde{c}
      \frac{k^{|I|}}{k^{(1+\epsilon)|I|}},
  \end{equation}
  where $\tilde{c}$ is a constant.  On the other hand by the remark
  above $\rho_k(\frac{1}{s_k}\xi)$ has norm bound from below by some
  constant. Thus, $\|\rho_k(r_k(\xi))-\rho(r_k(\xi_H))\|_{\tmop{End}(V_k)}
  \to 0$ as $k\to\infty$.
\end{proof}

Analogously, one would like to treat the case of ``underscaling'',
i.e., if $s_k$ grows slower than $k^{1-\epsilon}$. In that case we
would expect the limit of the nearest neighbour distribution to depend
only on the top homogeneous part of $\xi_H$. While this is the case in
all known examples, we can give no proof here. The problem being that
a lower bound for the growth of the norm in the top homogeneous part
is needed. If the norms of the top homogeneous part behave
asymptotically like in (\ref{eq:norm-estimate}) then with an analogous
argument like in the proof above one can show convergence to the
nearest neighbour distribution of the top homogeneous part.

\subsection{A remark on the commutativity}
\label{sec:remark-semicl-interp} Let us briefly discuss rescaling from the
semiclassical viewpoint, by pointing that the postulated rescaling ensures the
commutativity of the dynamical variables in the semiclassical limit. For simple
operators let's consider $r_k(\xi_1\xi_2)=\frac{1}{s_k^2}\xi_1\xi_2=
\frac{1}{s_k^2}\xi_2\xi_1+\frac{1}{s_k^2}[\xi_1,\xi_2]$.  As the commutator is
a linear combination of $\xi_j$, the respective term will vanish as $k$ goes to
infinity. A similar reasoning can be applied to arbitrary homogenous
polynomials in $\xi_j$. So, in this narrow sense the amount of commutativity is
increased, although the operators themselves are non-commutative. We like to
think of this as a sign of the semiclassical nature of the limit.

\section{The Lipkin-Hamiltonian}
\label{sec:lipkin-hamiltonian}

In this section we will make some remarks about more complicated
operators on $SU_3$, i.e.\ simple "polynomials" in the universal
enveloping algebra of $\mathfrak{sl}(3,\mathbb{C})$.

Recall that the fundamental irreducible representations $\rho_k$ of
the group $SU_n$ are given as the natural representations of $SU_n$ on
the vector spaces
\begin{equation}
  \bigwedge_{i=1}^k \mathbb{C}^n \text{ for }k=1,\dots,n-1.
\end{equation}
By the theorem of Peter and Weyl every irreducible
representation is realized as a subrepresentation of the natural
$SU_n$ representation on $L_2(SU_n)$. It is known, what the
generating functions for them are: let $g \in SU_n$ be the matrix
\begin{equation}
  \left(
    \begin{matrix}
       g_{11} & \dots & g_{1n} \\
       \vdots & & \vdots \\
       g_{n1} & \dots & g_{nn}
    \end{matrix}
  \right),
\end{equation}
then the vector space of the irreducible representation with highest
weight $\lambda=\sum_{k=1}^{n-1} \lambda_k f_k$, where $f_k$ denotes
the highest weight of $\rho_k$ is given as the linear span of the
$k$-homogeneous polynomials in the minors of $g$ of the appropriate
degree taken from the first columns only. This means, we take the
homogeneous polynomials of degree $\lambda_1$ in the $1\times
1$-minors of the first column, the homogeneous polynomials of degree
$\lambda_2$ in the $2\times 2$-minors of the first two columns and so
forth, and then take every possible multiplicative combination of them.

For $SU_3$ we have to consider only the $1\times 1$-minors
\begin{equation}
   x_1 := g_{11}, x_2 := g_{12}, x_3 := g_{13}
\end{equation}
and the $2\times 2$-minors
\begin{equation}
  y_1 := g_{21}g_{32}-g_{31}g_{22}, y_2 := g_{31}g_{12} - g_{11}g_{32},
  y_3 := g_{11}g_{22}-g_{21}g_{12}.
\end{equation}
A generating set for the representation subspace of $L_2(SU_3)$
corresponding to the highest weight $\lambda=(\lambda_1,\lambda_2)$ is
then generated by the polynomials in the $x_i$ and $y_j$ which are
homogeneous of degree $\lambda_1$ in the $x_i$ and homogeneous of
degree $\lambda_2$ in the $y_j$.  Unfortunately, these are not always
linearly independent due to the relation
\begin{equation}
  x_1 y_1 + x_2 y_2 + x_3 y_3  = 0.
\end{equation}

It can be shown, that this is the only relation among the variables, i.e., the
irreducible representations are given as the bi-graded subrepresentations of
the representation on the $\mathbb{C}$-algebra
\begin{equation}
  R:=\mathbb{C}[x_1,x_2,x_3,y_1,y_2,y_3]/\langle x_1 y_1 + x_2 y_2 +
  x_3 y_3\rangle.
\end{equation}

The necessary details for calculating these relations for arbitary
$SU_n$ can be found in \cite{millersturmfels} and
\cite{doubiletrota}. In the first book there is actually a nice
interpretation in terms of Gr\"obner bases.

It was shown in the last section that elements from $\mathfrak{su}_3$
will have spectral statistics converging to the Dirac measure in the
interior of the Weyl chamber. We want to study a little more
complicated object, called the Lipkin-Hamiltonian in nuclear physics
(cf.\cite{haakekus}). It is an abstractly hermitian operator in
$\mathcal{T}(\mathfrak{sl}_3(\mathbb{C}))$ given by
\begin{equation}
  \label{eq:lipkin-hamiltonian}
  \xi_{\tmop{Lipkin}} = a T_3 + b \sum_{i\neq j} S_{ij}^2,
\end{equation}
where $T_3=diag(1,0,-1)$ and $S_{ij}$ are the matrices with $1$ at
position $(i,j)$ and $0$ elsewhere and $a,b$ denote some real
constants. For $b=0$, we already know that the spectral statitstics in
the interior of the Weyl chamber converge to zero. Before we go any
further, we will first calculate, how these operators act.

We start with the example of $S_{12}$. It acts on a vector $v$ by
$\left. \frac{d}{dt}\right|_{t=0} exp(S_{12}t).v$. Since we are
acting on functions, the action on the argument is given by
multiplication with the inverse.
\begin{equation}
   exp(S_{12}t)^{-1} \left(
    \begin{matrix}
     g_{11} & g_{12} & g_{13} \\
     g_{21} & g_{22} & g_{23} \\
     g_{31} & g_{32} & g_{33}
    \end{matrix}
   \right) = \left(
    \begin{matrix}
     g_{11}-tg_{21} & g_{12}-tg_{22} & g_{13}-tg_{23} \\
     g_{21} & g_{22} & g_{23} \\
     g_{31} & g_{32} & g_{33}
     \end{matrix}
   \right).
\end{equation}
Thus, we get the following substitution rules for the arguments $x_i$
and $y_i$:
\begin{equation}
  \begin{aligned}
    x_1 = g_{11} & \mapsto g_{11}-tg_{21} = x_1-tx_2 \\
    x_2 = g_{21} & \mapsto x_2 \\
    x_3 = g_{31} & \mapsto x_3 \\
    y_1 = g_{21}g_{32}-g_{31}g_{22} & \mapsto y_1 \\
    y_2 = g_{31}g_{12}-g_{11}g_{32} & \mapsto g_{31}(g_{12}-tg_{12})-(g_{11}-tg_{21})g_{32} = y_2+ty_1 \\
    y_3 = g_{21}g_{32}-g_{31}g_{22} & \mapsto (g_{11}-tg_{21})g_{22}-g_{21}(g_{12}-tg_{22}) = y_3 \\
  \end{aligned}
\end{equation}
Applying this to the basis elements
$x_1^{a_1}x_2^{a_2}x_3^{a_3}y_1^{b_1}y_2^{b_2}y_3^{b_3}$ the action is
given by
\begin{equation}
  \begin{split}
    \left. \frac{d}{dt}\right|_{t=0} exp(S_{12}t).x_1^{a_1}x_2^{a_2}x_3^{a_3}y_1^{b_1}y_2^{b_2}y_3^{b_3}
    & = -a_1 x_1^{a_1-1}x_2^{a_2+1}x_3^{a_3}y_1^{b_1}y_2^{b_2}y_3^{b_3} \\
    & +  b_2
     x_1^{a_1}x_2^{a_2}x_3^{a_3}y_1^{b_1+1}y_2^{b_2-1}y_3^{b_3}
  \end{split}
\end{equation}

Thus, the $S_{12}$ action is given by the action of the differential
operator
\begin{equation}
   -x_2\frac{\partial}{\partial x_1} + y_1\frac{\partial}{\partial y_2}
\end{equation}
on the bi-homogeneous polynomials. In fact, analogously we can
calculate that the action of $S_{ij}$ is given by
\begin{equation}
  -x_j\frac{\partial}{\partial x_i} + y_i\frac{\partial}{\partial y_j}.
\end{equation}
The actions of $T_3$ corresponds to the differential operator
\begin{equation}
  -x_1\frac{\partial}{\partial x_1} +x_3\frac{\partial}{\partial
    x_3} + y_1\frac{\partial}{\partial y_1} - y_3\frac{\partial}{\partial y_3}.
\end{equation}

\begin{remark}
  The reader should be aware of the fact that in the chosen basis
  operators which are abstract hermitian, e.g., $S_{12}+S_{21}$, are not
  represented as hermitian matrices.
\end{remark}

We begin our discussion by showing that $T_3$ has the Dirac measure as limiting
spectral statistic in any case. Recall, that we proved this only for the
interior of the Weyl chamber in Section \ref{sec:simple-operators}.

\begin{lemma}
  Let $\lambda$ be the highest weight of a non-trivial irreducible
  representation of $SU_3$ and assume that $\lambda$ lies on the
  border of the Weyl chamber of $SU_3$, i.e.\ $\lambda=(\lambda_1,0)$
  or $\lambda=(0,\lambda_2)$ and $\lambda\neq(0,0)$.

  Then the spectral statistics of the operator
  $T=diag(1,0,-1)\in\mathfrak{sl}_3(\mathbb{C})$ along the irreducible
  representations $\rho_{\ast,m\lambda}$ converge to the Dirac measure
  for $m\to\infty$.
\end{lemma}

\begin{proof}
  For this we consider only the case $\lambda_2=0$, the other case
  being analogous.

  The element $T$ operates on a polynomial $x_1^{a_1} x_2^{a_2}
  x_3^{\lambda_1-a_1-a_2}$ by
  \begin{equation}
    T.x_1^{a_1} x_2^{a_2} x_3^{\lambda_1-a_1-a_2}=(-a_1+\lambda_1-a_1-a_2) x_1^{a_1} x_2^{a_2} x_3^{\lambda_1-a_1-a_2}.
  \end{equation}
  So the eigenvalues of $T$ are given by $-2a_1-a_2+\lambda_1$ with $0
  \leq a_1,a_2$ and $a_1+a_2\leq\lambda_1$. We see for example that the
  zero eigenvalue has increasing multiplicity if $\lambda_1$ increases,
  because
  \begin{equation}
    0 = 2\cdot 0 + \lambda_1 -\lambda_1 = 2\cdot 1 + (\lambda_1-2) - \lambda_1 = 2\cdot 2+ (\lambda_1-4) -
    \lambda_1 = \dots
  \end{equation}

  Let us count the number of different eigenvalues now. This number is
  the number the fibers of the map $(a_1,a_2)\mapsto 2a_1+a_2$, which
  is $2\lambda_1-1$. To see this, just note that $(a_1,a_2)$ is in the
  same fiber as $(a_1+1,a_2-2)$. So every fiber can be represented by
  an element of the form $(0,a_2)$ or $(a_1,\lambda_1)$ and these are
  exactly $2\lambda_1-1$ elements.

  But the dimension of the representation on the space on homogeneous
  polynomials of degree $\lambda_1$ is
  $\frac{1}{2}\lambda_1(\lambda_1-1)$. Thus the quotient of the number
  of eigenvalues and the dimension is
  \begin{equation}
    \frac{2\lambda_1-1}{\frac{1}{2}\lambda_1(\lambda_1-1)}= \frac{4}{\lambda_1-1}-\frac{2}{\lambda_1(\lambda_1-1)}
  \end{equation}
  and converges to zero if $\lambda_1\to\infty$. Thus, we have proved
  the lemma.
\end{proof}

Our basic input for the Lipkin Hamiltonian is the explicit action of
the $S_{ij}^2$. These act by sparse matrices as we are going to
discuss now.

Analogously, the action of other $S_{ij}$ can be calculated. We give
the action in a tabular form here, where we just write
$[\dots]:=[a_1,a_2,a_3,b_1,b_2,b_3]$ instead of the rather clumsy
$x_1^{a_1}x_2^{a_2}x_3^{a_3}y_1^{b_1}y_2^{b_2}y_3^{b_3}$.
\begin{equation}
  \begin{split}
    S_{12}&: [\dots] \mapsto -a_1[a_1-1,a_2+1,a_3,b_1,b_2,b_3] + b_2[a_1,a_2,a_3,b_1+1,b_2-1,b_3] \\
    S_{13}&: [\dots] \mapsto -a_1[a_1-1,a_2,a_3+1,b_1,b_2,b_3] + b_3[a_1,a_2,a_3,b_1+1,b_2,b_3-1] \\
    S_{21}&: [\dots] \mapsto -a_2[a_1+1,a_2-1,a_3,b_1,b_2,b_3] + b_1[a_1,a_2,a_3,b_1-1,b_2+1,b_3]\\
    S_{23}&: [\dots] \mapsto -a_2[a_1,a_2-1,a_3+1,b_1,b_2,b_3] + b_3[a_1,a_2,a_3,b_1,b_2+1,b_3-1]\\
    S_{31}&: [\dots] \mapsto -a_3[a_1+1,a_2,a_3-1,b_1,b_2,b_3] + b_1[a_1,a_2,a_3,b_1-1,b_2,b_3+1]\\
    S_{32}&: [\dots] \mapsto -a_3[a_1,a_2+1,a_3-1,b_1,b_2,b_3] + b_2[a_1,a_2,a_3,b_1,b_2-1,b_3+1]
  \end{split}
\end{equation}

Applying every $S_{ij}$ twice we get
\begin{equation}
 \begin{aligned}
    S_{12}^2: [\dots] \mapsto &   a_1(a_1-1)\ [a_1-2,a_2+2,a_3,b_1,b_2,b_3] \\
                              & + b_2(b_2-1)\ [a_1,a_2,a_3,b_1+2,b_2-2,b_3] \\
                              & -2a_1b_2\     [a_1-1,a_2+1,a_3,b_1+1,b_2-1,b_3] \\
    S_{13}^2: [\dots] \mapsto &   a_1(a_1-1)\ [a_1-2,a_2,a_3+2,b_1,b_2,b_3] \\
                              & + b_3(b_3-1)\ [a_1,a_2,a_3,b_1+2,b_2,b_3-2] \\
                              & -2a_1b_3\     [a_1-1,a_2,a_3+1,b_1+1,b_2,b_3-1] \\
    S_{21}^2: [\dots] \mapsto &   a_2(a_2-1)\ [a_1+2,a_2-2,a_3,b_1,b_2,b_3] \\
                              &+  b_1(b_1-1)\ [a_1,a_2,a_3,b_1-2,b_2+2,b_3]\\
                              & -2a_2b_1\     [a_1+1,a_2-1,a_3,b_1-1,b_2+1,b_3]\\
                              & \vdots & \\
%    S_{23}^2: [\dots] \mapsto &-a_2&[a_1,a_2-1,a_3+1,b_1,b_2,b_3] + b_3[a_1,a_2,a_3,b_1,b_2+1,b_3-1]\\
%    S_{31}^2: [\dots] \mapsto &-a_3&[a_1+1,a_2,a_3-1,b_1,b_2,b_3] + b_1[a_1,a_2,a_3,b_1-1,b_2,b_3+1]\\
%    S_{32}^2: [\dots] \mapsto &-a_3&[a_1,a_2+1,a_3-1,b_1,b_2,b_3] + b_2[a_1,a_2,a_3,b_1,b_2-1,b_3+1]
  \end{aligned}
\end{equation}

From this we can read of the matrix representation of the $S_{ij}^2$
on the vector space of bi-graded polynomials of bidegree
$(\lambda_1,\lambda_2)$ in $R$, where we choose the representatives
$[a_1,a_2,a_3,b_1,b_2,b_3]$ with $a_3b_3=0$, i.e.\ we replace every
$x_3y_3$ by $-x_1y_1-x_2y_2$. We see immediately that $S_{12}^2$ and
$S_{21}^2$ do not increase $a_3$ or $b_3$, so the matrix
representation of $S_{12}^2$ and $S_{21}^2$ has at most 3 entries not
equal to zero in each column. The other $S_{ij}$ have two summands
which may increase either $a_3$ or $b_3$, so they have at most 5
non-zero entries in each column. Thus, the sum $\sum_{i\neq
  j}S_{ij}^2$ has at most $2\cdot 3+4\cdot 5=26$ non-zero entries in
each column, regardless of the bi-degree $(\lambda_1,\lambda_2)$. Each
of these entries has absolute value bounded by
$\max\{\lambda_1,\lambda_2\}^2$.

So, the it possible to compute the Lipkin Hamiltonian numerically and get 
graphical representations as in the following figure:
\begin{center}
\includegraphics[width=10cm, angle=270]{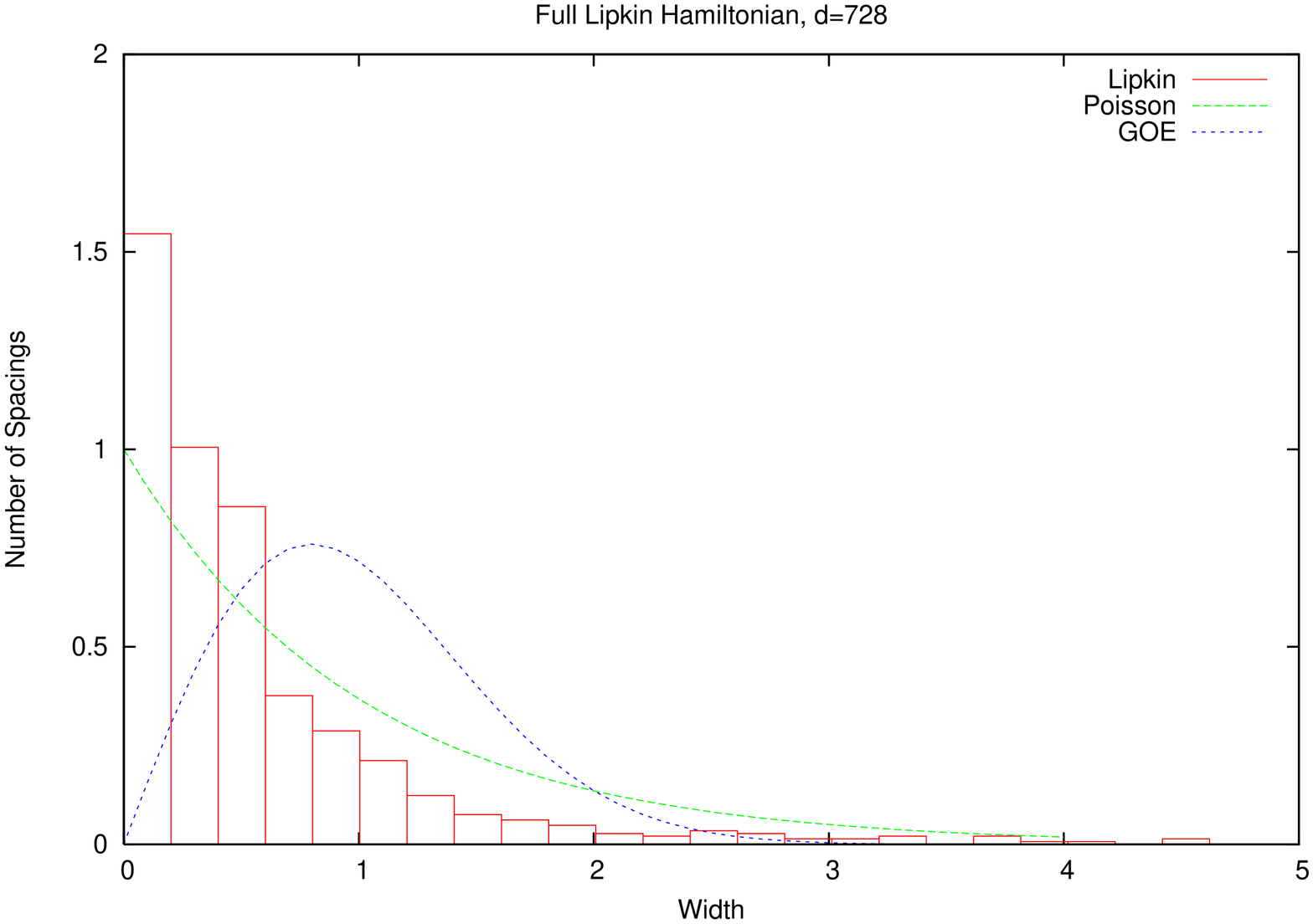}
\end{center}
Here, the nearest neighbor statistics are drawn for the Lipkin
Hamiltonian in the irreducible representation corresponding to
$\lambda_1=\lambda_2=8$, which is 728 dimensional.

\section{Acknowledgments}
The support by SFB/TR12 `Symmetries and Universality in Mesoscopic Systems'
program of the Deutsche Forschungsgemeischaft and Polish MNiSW grant No
1P03B04226 is gratefully acknowledged.

%\bibliography{Literatur}
%\bibliographystyle{alpha}

\end{document}